\documentclass[12pt]{elsarticle}
\pdfoutput=1
\usepackage{amssymb}
\usepackage{amsmath}
\usepackage{amsfonts}
\usepackage{graphicx}
\usepackage{xcolor}

\journal{Computers and Mathematics with Applications}

\begin{document}

\begin{frontmatter}

\title{Domain growth in cholesteric blue phases: hybrid lattice Boltzmann
simulations}

%\tightenlines

\author{O. Henrich, D. Marenduzzo, K. Stratford, M. E. Cates}

\address{SUPA, School of Physics, University of Edinburgh, Mayfield Road
Edinburgh EH9 3JZ, Scotland}

\begin{abstract}
Here we review a hybrid lattice Boltzmann algorithm to solve the 
equations of motion of cholesteric liquid crystals. The method consists in 
coupling a lattice Boltzmann solver for the Navier-Stokes equation to a 
finite difference method to solve the dynamical equations governing the 
evolution of the liquid crystalline order parameter. We apply this method to 
study the growth of cholesteric blue phase domains, within a cholesteric
phase. We focus on the growth of blue phase II and on a thin slab geometry 
in which the domain wall is flat. Our results show that, depending on
the chirality, the growing blue phase is either BPII with no or
few defects,
or another structure with hexagonal ordering. We hope that our simulations
will spur further experimental investigations on quenches in micron-size blue 
phase samples. The computational size that our hybrid lattice Boltzmann 
scheme can handle suggest that large scale simulations of new generation of 
blue phase liquid crystal device are within reach.
\end{abstract}

\begin{keyword}
lattice Boltzmann simulations \sep liquid crystals \sep cholesterics
\sep blue phases

% 61.30.Mp Blue phases and other defect-phases 
%64.70.mf Theory and modeling of specific liquid crystal transitions, including %computer simulation
%61.30.-v Liquid crystals (for phase transitions in liquid crystals, see 64.70.M-; for liquid crystals as dielectric materials, see 77.84.Nh; for liquid crystals as optical materials, see 42.70.Df; for liquid crystal devices, see 42.79.Kr)

\PACS 64.70.mf \sep 61.30.Mp \sep 61.30.-v

% MSC codes here, in the form: \MSC code \sep code
% or \MSC[2008] code \sep code (2000 is the default)
% 76M20 Finite difference methods 
% 76M25 Other numerical methods 

\MSC 76M20 \sep 76M25 \sep 82-08

\end{keyword}

\end{frontmatter}

\section{Introduction}

\noindent The lattice Boltzmann (LB) algorithm~\cite{Succi,Boghosian} is a powerful 
method to solve the Navier-Stokes equations of ideal and complex fluids, 
which, due to its conceptual simplicity and ease to code, provides an 
attractive alternative to other methods more commonly adopted in
computational fluids dynamics, such as e.g. finite elements algorithms.
In the last couple of decades, the LB method has evolved 
to a powerful tool to study a variety of problems in 
complex fluids \cite{bijel,bijel_review,Swift96,Wagner98,Gonnella97,Xu06,Xu06HLB,Denniston04,Marenduzzo07,Cates08,Kevin1,Kevin2}. In this respect, 
particularly successful avenues have been, among others, the study of
binary fluids and of liquid crystals.
  
The original LB method to solve the hydrodynamic equations of motion
of complex fluids typically consisted of introducing extra distribution
function, which evolved via an LB dynamics determined by an appropriate
collision kernel. 
These distribution functions control the dynamics of conserved or
non-converved order parameters which is coupled to the fluid
momentum, itself governed by the primary LB distribution function.
The Chapman-Enskog expansion of such dynamics 
then gave back the continuum hydrodynamic equations for the coupled
system. However, on top
of introducing some systematic, albeit small, errors, this ``full LB''
approach had the drawback of requiring a large memory to store all the 
distribution functions for the whole lattice. This turned out to be particularly
severe for the case of liquid crystal, where the order parameter (in the
Beris-Edwards model \cite{Beris}) is a traceless and symmetric tensor, and therefore
required the introduction of five extra sets of distribution functions. This leads to high memory requirements which ultimately limits the size of 
the lattice to be studied. The new generation of LB studies for complex fluids
has a growing need to be able to cope with larger and larger systems,
to make best use of the potentially very good scalability of parallel LB codes.
As a result, new hybrid algorithms have been coded and deployed, for both
binary fluids and liquid crystals \cite{Xu06HLB,Marenduzzo07,Ronojoy}, where 
the LB algorithm is dedicated to solve the (forced)
Navier-Stokes equation for momentum alone, and is
coupled to a standard efficient finite-difference solver (e.g. a predictor
corrector) for the order parameter dynamics. Hybrid LB simulations of 
binary fluids~\cite{Xu06HLB}
and of active gel and active liquid crystal 
hydrodynamics and rheology~\cite{Marenduzzo07,Cates08}
have validated the hybrid algorithm through
a comparison with full LB simulations, and have shown that this framework
is potentially very flexible and robust. 
Here, after briefly reviewing the equations of motion of liquid
crystal hydrodynamics and the hybrid LB approach which we use to solve them, 
we apply our algorithm to study the growth of domains in cholesteric blue 
phases (BPs).

BPs are a spectacular example of a soft solid, made up by a spontaneously 
occurring network of disclination lines in a cholesteric liquid 
crystal~\cite{Wright89}.
BPs appear very close to the transition between the isotropic and the
cholesteric phase.
In a liquid crystal in the cholesteric phase, the director field, which
quantifies the local direction of molecular ordering in the fluid, 
has the tendency to twist around an axis, which is that of 
the cholesteric helix.
Close to the transition to the isotropic phase, a simple helical order
is frustrated and it is more advantageous for the director field to
rotate in a helical fashion around any axis perpendicular to a line - this 
director field pattern was named a "double twist cylinder". Mathematically, it 
is impossible to patch up different double twist cylinders without creating 
defects in between, and this is what gives rise to the disclination network 
making up BPs.

Without an electric field, three different kinds of BPs are found.
BPI and BPII are cubic phases characterised by a regular disclination
line lattice: in BPI the double twist cylinders are arranged in a simple
cubic lattice, whereas in BPII they form a body-centred cubic lattice.
BPIII, or the ``blue fog'', is not a cubic phase, and its structure is
not fully understood to date: the dominating view is that it is a structure
made up by double twisted cylinders arranged in an irregular 
way~\cite{Wright89}. 

As the period of the BP network and the timescales of response
to an applied perturbation are in the sub-$\mu$m and sub-$\mu$s
range respectively, these soft materials are promising candidates for
tunable photonic crystals and a new generation of fast liquid crystal devices.
Recent experiments \cite{Coles05} managed to stabilize BPs over 
a strikingly wide temperature range of 50 K, compared to the preceedingly 
narrow temperature range of about 1 K, putting these technological
advances now within reach. However, in order for this exciting potential to be 
fully realised, our understanding of BPs needs to become as quantitative as 
the one we have for conventional nematic liquid crystal displays. 

From a more fundamental physics points of view, BPs are remarkable 
materials. A series of very interesting 
experiments~\cite{Pieranski85,Pieranski00,Pieranski08} have shown 
for instance that BP 
droplets facet when they nucleate and grow inside another 
isotropic fluid. BP droplets, like other fluids, may also wet a surface, but 
due to their elastic properties their surface shows steps and may
reconstruct, like that of a solid. According to the current
understanding, the underlying disclination meshwork
is primarily cause of this highly non-trivial phenomenology,
as it behaves like an elastic network.

Existing simulations of BPs~\cite{Dupuis05a,Alexander06,Dupuis05b,Alexander08} 
have significantly extended our quantitative
understanding of their physics, which first rested on semi-analytical
approximations (see e.g. ~\cite{Grebel}).
For instance, obtaining the shape of the phase
diagram, with BPI and BPII appearing in the experimentally observed order 
upon increasing the chirality~\cite{Dupuis05a,Alexander06}, as well as 
understanding the presence of anomalous electrostriction in BPI under an 
electric field~\cite{Alexander08}, were
only possible with extensive simulations of these structures. 
However, these simulations have thus far been
limited to one unit cell, within which 
several disclination cores are present and
require a fine enough discretisation to be correctly resolved.
By necessity, this approach leaves out a number of physically relevant
supra-unit cell
phenomena such as the possibility of large scale lattice 
reconstructions (perhaps induced by shear or by an applied field), 
and the appearance of defects or spatial inhomogeneities in the BP lattice. 
Furthermore, the length scale covered by simulations is much smaller 
than the ones relevant for either the domain growth experiments or BP devices.
As a first step to model more realistic situations, we present here
supra-unit cell large scale 3D simulations of the domain growth of BP 
domains in cholesteric and isotropic liquid crystals. 

Our hybrid lattice Boltzmann results suggest that the physics of BP
domain growth is highly non-trivial. Focussing on a thin slab geometry, 
we observe below qualitatively different growth dynamics of BP domains inside
cholesterics, for different values of the chirality. For large values of this
parameter, we find that BPII domains growing inside a cholesteric evolve into 
a different disclination network, which possesses local hexagonal symmetry.
Further simulations in an isotropic fluid show the same transition,
although the growth kinetics of the new BP phase is different.
To aid comparison with potential future experiments, we also visualize the 
disclination network and simulate 
the appearance of the sample under polarized light. 
The feasibility of such large scale simulations is 
promising in view of further future applications, e.g. to device modelling. 

\section{Model and methods}

\subsection{Equations of motion}

\noindent Following the Beris-Edwards model for liquid crystal hydrodynamics 
\cite{Beris}, we describe the BPs by a traceless, symmetric, second rank 
order parameter tensor, $\mathbf{Q}$. The equilibrium
thermodynamic properties are determined by a Landau - de Gennes free energy
${\cal F}$, whose density ${f}$ is,
\begin{eqnarray}
f  &=& \tfrac{A_0}{2} \bigl( 1 - \tfrac{\gamma}{3} \bigr) Q_{\alpha \beta}^2 
           - \tfrac{A_0 \gamma}{3} Q_{\alpha \beta}
     Q_{\beta \gamma}Q_{\gamma \alpha}
           + \tfrac {A_0 \gamma}{4} (Q_{\alpha \beta}^2)^2 \nonumber\\ 
	 &  &+ \tfrac{K}{2} 
           \bigl( \epsilon_{\alpha \gamma \delta} \partial_{\gamma} 
Q_{\delta \beta} 
           + 2q_0 Q_{\alpha \beta} \bigr)^2 .
\label{free_energy}
\end{eqnarray}
(Notice that in our notation Greek indices denote Cartesian components
and summation over repeated indices is implied.)
In Eq. \ref{free_energy} $A_0$ is a constant, $K$ is an elastic constant (we 
adopt the one-constant-approximation), $q_0$ is $2\pi/p$, where $p$ is
the pitch of the 
cholesteric helix, while $\gamma$ is a control parameter linked to
the temperature for 
thermotropic liquid crystals, or concentration for lyotropic ones. 
Increasing $\gamma$ leads to an increase in the average magnitude of order.
The local magnitude of order, $q(\vec r)$, is
the largest eigenvalue of ${\bf Q}$.
This quantity is also used to identify disclination lines: when the
local magnitude of order falls below a predetermined threshold, we identify
that lattice point as constituting part of a disclination. This prescription
is very easy to implement and allows an accurate determination of the
disclinations. Changing the numerical value of the threshold generally
leads to a change in the thickness of the disclination tubes. For our
study, we have typically chosen a threshold of $q=0.19$ for defect/disclination
identification. 
The Beris-Edwards equations for the evolution 
of the $\mathbf{Q}$-tensor, which we aim to solve, are~\cite{Beris}
\begin{equation}
D_t \mathbf{Q} 
= \Gamma  \Bigl( \tfrac{-\delta {\cal F}}{\delta \mathbf{Q}} + \tfrac{1}{3}\, 
\text{Tr} \Bigl( \tfrac{\delta {\cal F}}{\delta \mathbf{Q}} \Bigr) \mathbf{I} \Bigr) \equiv \Gamma\, {\mathbf H}  .
\label{Qevolution}
\end{equation} 
Here $\Gamma$ is a collective rotational diffusion constant, while $\frac{\delta}{\delta {\bf Q}}$ indicates 
the functional derivative with respect to the tensor order parameter. $\text{Tr}$
stands for trace, and $D_t$ is the 
material derivative for rod-like molecules~\cite{Beris}:
\begin{eqnarray}
\label{mat_der}
D_t{\bf Q} & \equiv & (\partial_t+{\vec u}\cdot{\bf \nabla}){\bf Q}-{\bf S}({\bf W},{\bf
  Q}) \\
{\bf S}({\bf W},{\bf Q})
& \equiv &(\xi{\bf D}
+{\bf \omega})({\bf Q}+{\bf I}/3)\\ \nonumber
&+& ({\bf Q}+
{\bf I}/3) (\xi{\bf D}-{\bf \omega})
- 2\xi({\bf Q}+{\bf I}/3){\mbox{Tr}}({\bf Q}{\bf W}),
\end{eqnarray}
where ${\bf D}$ and ${\bf \omega}$ are the
symmetric and the anti-symmetric part respectively of the
velocity gradient tensor $W_{\alpha\beta}=\partial_\beta
u_\alpha$ \cite{Beris,Denniston04}, $\vec u$ being the velocity field. 

Eq. \ref{Qevolution} may be mathematically derived from an underlying
microscopic model via a Poisson brackets approach \cite{Beris}. Physically, 
this equation means that the system tries to evolve, in the absence of any flow
or backflow ($\vec u=0$), so as to minimise its free energy. This is
ensured by the presence of the molecular field,  $\mathbf{H}$.
The presence of
a non-zero $\vec u$ requires one to substitute the usual
partial derivative $\partial_t$ with the material derivative $D_t$,
which describes advection by the fluid velocity (the term 
${\vec u}\cdot{\bf \nabla}{\bf Q}$), and also includes a further 
coupling ${\bf S}({\bf W},{\bf Q})$ between the velocity gradient tensor 
and the order parameter, which arises due to the tensorial nature of the 
latter. Physically, ${\bf S}({\bf W},{\bf Q})$ appears
because the order parameter distribution can be both rotated and
stretched by flow gradients \cite{Beris}. Mathematically, 
${\bf S}({\bf W},{\bf Q})$ contains a (phenomenological)
mixture of upper and lower convected derivatives. This is similar to
what is done in other hydrodynamic equations for polymeric fluids, such 
as the Johnson-Segalman model \cite{Larson}.
The term $\xi$, which may be also viewed as a parameter controlling the
weights of the lower and upper convected derivatives,
is related to the aspect ratio of the molecules. 
This parameter controls whether the director field is flow aligning
in shear flow ($\xi \ge 0.6$), creating a stable response, 
as opposed to flow tumbling, which gives an unsteady response
($\xi<0.6$).

The fluid velocity, $\vec u$, obeys the continuity equation
of an incompressible fluid, and
the Navier-Stokes equation,
\begin{eqnarray}\label{navierstokes}
\rho(\partial_t+ u_\beta \partial_\beta)
u_\alpha & = & \partial_\beta (\Pi_{\alpha\beta})+
\eta \partial_\beta(\partial_\alpha
u_\beta + \partial_\beta u_\alpha\\ \nonumber
& + & (1-3\partial_\rho
P_{0}) \partial_\gamma u_\gamma\delta_{\alpha\beta}),
\end{eqnarray}
where $\rho$ is the fluid density, $\eta$ is an isotropic
viscosity. The stress tensor $\Pi_{\alpha\beta}$ consists of a 
symmetric part $\sigma_{\alpha\beta}$ and an antisymmetric contribution $\tau_{\alpha\beta}$, 
\begin{eqnarray}
\Pi_{\alpha\beta}&=&\sigma_{\alpha\beta}+\tau_{\alpha\beta}\\
\sigma_{\alpha\beta}&=&-P_0 \delta_{\alpha \beta} +2\xi
(Q_{\alpha\beta}+\tfrac{1}{3}\delta_{\alpha\beta})Q_{\gamma\epsilon}
H_{\gamma\epsilon}\\\nonumber
&-&\xi H_{\alpha\gamma}(Q_{\gamma\beta}+\tfrac{1}{3}
\delta_{\gamma\beta}) - \xi (Q_{\alpha\gamma}+\tfrac{1}{3}
\delta_{\alpha\gamma})H_{\gamma\beta} \\ \nonumber
&-& \partial_\alpha Q_{\gamma\nu} \frac{\partial
 f}{\partial Q_{\gamma\nu,\beta}}\\
\tau_{\alpha\beta}&=& Q_{\alpha \gamma} H_{\gamma \beta} -H_{\alpha
 \gamma}Q_{\gamma \beta} ,
\label{BEstress}
\end{eqnarray}
where $P_0=\rho T-f$ is an isotropic pressure \cite{Beris} and $Q_{\gamma\nu,\beta}=\partial_{\beta} Q_{\gamma\nu}$. 

In our simulations, the coupling to hydrodynamics via a non-trivial
pressure tensor can be switched off, essentially by imposing a constant
zero velocity profile. In this way the effects of flow and backflow
may be unambiguously pinpointed. In our case, backflow 
does not dramatically
modify the kinetic pathway through which BP domains grow, but it renders
the dynamics faster. Similar effects have been seen 
on the switching dynamics and on the rheological properties of BPs
\cite{Dupuis05b,Alexander08}.

The equilibrium phase diagrams of BPs are commonly expressed as a function of 
the chirality $\kappa$ and reduced temperature $\tau$, defined \cite{Dupuis05a,Alexander06,Grebel} as
\begin{eqnarray}
\kappa&=&\sqrt{\frac{108 L_1 q_0^2}{A_0 \gamma}}\\
\tau&=&\frac{27(1-\gamma/3)}{\gamma}.
\end{eqnarray} 
We selected $\tau=0$ and 
varied $\kappa$ in our simulations (reported in the Results section).
For the other kinetic parameters, we chose $\Gamma=0.3$, $\xi=0.7$ 
(hence a flow aligning BP) and $\eta=5/3$ 
(these choices are
suggested by previous experience in blue phase numerics~\cite{Dupuis05b,Alexander08}). Time and space is measured in simulation units in what follows.

\subsection{The hybrid lattice Boltzmann algorithm}

\noindent In order to solve Eqs. \ref{Qevolution} and \ref{navierstokes},
as mentioned in the Introduction, we use here a hybrid lattice 
Boltzmann algorithm. The idea of the algorithm is simple.
We observe that the coupling between the velocity field and the order
parameter equation (see Eq. \ref{Qevolution}) is via the material derivative
term, which requires both the velocity and the velocity gradient fields.
On the other hand, the order parameter field affects the Navier-Stokes
equation through the pressure tensor $\Pi_{\alpha\beta}$.
Our hybrid lattice Boltzmann approach consists in solving 
Eq. \ref{Qevolution} via a finite-difference predictor-corrector 
algorithm, while the LB algorithm is devoted to the integration
of Eq. \ref{navierstokes}. The order parameter and velocity fields
are (sequentially) updated at every time step via these algorithms.
The LB step requires as an input the order parameter field (hence the
pressure tensor and its divergence), which is provided by the finite
difference solver. On the other hand, the LB algorithm updates the
velocity field which is then in turn required by the finite difference
scheme to further evolve the dynamics of the order parameter.
Due to the limited required coupling between LB and finite difference 
algorithms, this hybrid approach could be generalised to a variety of
hydrodynamic equations of motion involving the dynamics of an order
parameter besides the Navier-Stokes 
equations~\cite{Xu06HLB,Marenduzzo07,Ronojoy}. As mentioned above, a 
similar hybrid approach has indeed already been successfully used for
a binary fluid as well \cite{Xu06HLB,Ronojoy}.

With respect to a full LB approach
\cite{Denniston01,Denniston04}, the primary advantage of this hybrid method is
that it will allow simulations of larger systems as it involves
consistently smaller memory requirements. Indeed, while in a full
LB treatment one has to store 6 sets of 15 distribution functions
at any lattice point (if we choose the 3DQ15 velocity vector
lattice \cite{Succi} as we do here), in the
hybrid approach just one set of distribution functions plus the
five independent components of the ${\bf Q}$ tensor are needed.
The hybrid method should also be numerically more stable since it avoids
the error term arising in the Chapman-Enskog expansion used to
connect the LB model to the order parameter evolution equation in
the continuum limit \cite{Denniston01}. Finally, it should be
possible to include noise in a conceptually simpler way as one can
in principle couple a noisy LB dynamics, for which precise prescriptions
exist \cite{LBnoise}, for the Navier-Stokes
equation with a Langevin dynamics for the order parameter(s) evolution.
In the present work, however, we neglect noise. Thus our treatment
couples a mean-field dynamics of the
cholesteric order to a noiseless classical fluid.

Let us describe first how to integrate the Navier-Stokes equations
with a slightly modified standard Lattice Boltzmann algorithm \cite{Denniston01}.  This is defined  in terms of
a single set of partial distribution functions, the scalars $f_i
(\vec{x})$, that sum on each lattice site $\vec{x}$ to give the
density. (Note that $f$ without Latin index is the free energy density from Eq. \ref{free_energy}.)
 Each $f_i$ is associated with a lattice
vector ${\vec e}_i$ \cite{Denniston01,Denniston04}.  
We choose a 15-velocity model on
the cubic lattice with lattice vectors:
\begin{eqnarray}
\vec {e}_{i}^{(0)}&=& (0,0,0)\\
\vec {e}_{i}^{(1)}&=&(\pm 1,0,0),(0,\pm 1,0), (0,0,\pm 1)\\
\vec {e}_{i}^{(2)}&=&(\pm 1, \pm 1, \pm 1). \label{latvects}
\end{eqnarray}
The indices, $i$, are ordered so that $i=0$ corresponds to $\vec
{e}_{i}^{(0)}$, $i=1,\cdots, 6$ correspond to the $\vec
{e}_{i}^{(1)}$ set and $i=7,\cdots,14$ to the $\vec {e}_{i}^{(2)}$
set.

Physical variables are defined as moments of the distribution
functions:
\begin{equation}
\rho=\sum_i f_i, \qquad \rho u_\alpha = \sum_i f_i  e_{i\alpha}.
\label{eq1}
\end{equation}
The distribution functions evolve in a time step $\Delta t$
according to
\begin{equation}
f_i({\vec x}+{\vec e}_i \Delta t,t+\Delta t)-f_i({\vec x},t)=
\frac{\Delta t}{2} \left[{\cal C}_{fi}({\vec x},t,\left\{f_i
\right\})+ {\cal C}_{fi}({\vec x}+{\vec e}_i \Delta t,t+\Delta
t,\left\{f_i^*\right\})\right]. \label{eq2}
\end{equation}
This represents free streaming with velocity ${\vec e}_i$ followed
by a collision step which allows the distributions to relax
towards equilibrium. The $f_i^*$'s are first order approximations
to $f_i(\vec{x}+\vec{e}_i dt,t+dt)$, and they are obtained
by using $\Delta t\, {\cal C}_{{f}i}({\vec x},t,\left\{{f}_i
\right\})$ on the right hand side of Eq. (\ref{eq2}). Discretizing
in this way, which is similar to a predictor-corrector scheme, has
the advantages that lattice viscosity terms are eliminated to
second order and that the stability of the scheme is improved
\cite{Denniston01}.

The collision operators are taken to have the form of a single
relaxation time Boltzmann equation, together with a forcing term
\begin{equation}
{\cal C}_{fi}({\vec {x}},t,\left\{f_i \right\})=
-\frac{1}{\tau_f}(f_i({\vec {x}},t)-f_i^{eq}({\vec
{x}},t,\left\{f_i \right\})) +p_i({\vec {x}},t,\left\{f_i
\right\}), \label{eq4}
\end{equation}
The form of the equations of motion follow from the choice of the
moments of the equilibrium distributions $f^{eq}_i$ and the
driving terms $p_i$. Note that $f_i^{eq}$ is constrained by
\begin{eqnarray}\label{eq6a}
\sum_i f_i^{eq} = \rho,\qquad \sum_i f_i^{eq} e_{i \alpha} = \rho
u_{\alpha}, \qquad \sum_i f_i^{eq} e_{i\alpha}e_{i\beta} =
-\sigma_{\alpha\beta}+\rho u_\alpha u_\beta, \\
\sum_i f_i^{eq} e_{i\alpha}e_{i\beta}e_{i\gamma} =
\frac{\rho \tau_f}{3} \left(u_{\alpha}\delta_{\beta\gamma}+
u_{\beta}\delta_{\alpha\gamma}+u_{\gamma}\delta_{\alpha\beta}\right)
\label{eq6b}
\end{eqnarray}
where the zeroth and first moments are chosen to impose
conservation of mass and momentum. The second moment of $f^{eq}$
is determined by $\sigma_{\alpha\beta}$, which is the symmetric part
of the stress tensor $\Pi_{\alpha\beta}$, and does {\em not} include 
either the double
gradient term, $\partial_\alpha Q_{\gamma\nu}
{\partial f \over \partial Q_{\gamma\nu,\beta}}$,
or the $ f \delta_{\alpha\beta}$ contribution from the isotropic
pressure.
The constraint on the third moment is necessary to get an isotropic
Navier-Stokes equation via the Chapman-Enskog expansion of
Eq. \ref{eq2} (see e.g. \cite{Denniston01}).

The divergences of $\tau_{\alpha\beta}$ and of $\partial_\alpha Q_{\gamma\nu}
{\partial { f}\over \partial Q_{\gamma\nu,\beta}}$ instead 
enter effectively as a body force and constrain the first moment of
the driving terms $p_i$: this ensures that spurious velocities
are eliminated, or greatly reduced, as we show below. The constraints
on the $p_i$'s are~\cite{note_bodyforce}:
\begin{eqnarray}
\label{eq7a}
\sum_i p_i = 0, \quad \sum_i p_i e_{i\alpha} = \partial_\beta
\tau_{\alpha\beta}-\partial_\beta \left(\partial_\alpha
Q_{\gamma\nu} {\partial f \over \partial
Q_{\gamma\nu,\beta}}\right)+\partial_{\alpha}{f}\equiv b_{\alpha}, \\ 
\sum_i p_i e_{i\alpha}e_{i\beta} = 0,\quad 
\sum_i p_i e_{i\alpha}e_{i\beta}e_{i\gamma}=0.
\label{eq7b}
\end{eqnarray}
%We note that any additional body force, e.g. coming from an external
%pressure difference which would cause the BP to flow, would simply
%be added to the ``intrinsic'' body force $b_{\alpha}$. 
This scheme
was inspired by the one used in Ref. \cite{Wagner03} to reduce spurious
velocities in a full LB binary fluid simulation. 
In the case of liquid crystals spurious velocities are equally eliminated by
introducing the divergence of $g_{\alpha \beta}=-\partial_\alpha Q_{\gamma\nu}
{\partial  f \over \partial Q_{\gamma\nu,\beta}}+
f\delta_{\alpha\beta}.$ To see this, we transform 
$\partial_{\beta} g_{\alpha \beta}$ in the following way:
\begin{eqnarray}
& - & \partial_{\beta}\left(\partial_\alpha Q_{\gamma\nu}
{\partial  f\over \partial Q_{\gamma\nu,\beta}}
\right)+\partial_{\alpha} f =  \nonumber \\
& - &\partial_{\beta} 
\left( {\partial f\over \partial  Q_{\gamma\nu,\beta}}\right)
\partial_{\alpha} Q_{\gamma\nu,\beta}- 
{\partial f\over \partial  Q_{\gamma\nu,\beta}}
\partial_{\alpha} Q_{\gamma\nu,\beta}  \nonumber\\
&+& \left({\partial f \over{\partial Q_{\gamma\nu}}} + 
{\partial f \over {\partial  Q_{\gamma\nu,\beta}}}\right)
\partial_{\alpha} Q_{\gamma\nu} = 
-H_{\gamma\nu}\partial_{\alpha}Q_{\gamma\nu}\label{spurious},
\end{eqnarray}
where we introduced $H_{\gamma\nu}$ in the last line because 
the ${\mathbf Q}$ tensor is traceless and therefore $Q_{\gamma\gamma}=0$.
We see that if we directly insert Eq. \ref{spurious} as a body force
 this term vanishes in equilibrium as
it is proportional to the molecular field. In this way spurious
velocities, i.e. non-zero velocities in equilibrium due to the
different discretisation of the pressure tensor and the molecular
field, are in principle avoided. We have verified that this is the case
numerically.

Conditions Eqs. \ref{eq6a} -- \ref{eq7b} are satisfied by writing the
equilibrium distribution functions and forcing terms as polynomial
expansions in the velocity,
\begin{equation}
f_i^{eq}=A_s + B_s u_\alpha e_{i\alpha}+C_s u^2+D_\alpha u_\alpha
u_\beta e_{i\alpha}e_{i\beta}+E_{\alpha\beta}e_{i\alpha}e_{i\beta}
+ P_s  b_{\alpha}e_{i\alpha},
\label{feq}
\end{equation}
where $s\in \{0,1,2\}\Leftrightarrow {\vec e}_i\;^2 \in \{0,1,3\}$ identifies separate
coefficients for different square absolute values of the velocities.

The coefficients in the expansion are 
determined by the requirements that all the constraints 
enumerated above, Eqs. \ref{eq6a},\ref{eq6b},\ref{eq7a} and \ref{eq7b}, are
fulfilled.  A possible choice, which we adopt, is given below:

\begin{eqnarray}
&&A_2=\frac{\rho T}{10},\qquad A_1= A_2,\qquad A_0=\rho- 14 A_2,\nonumber \\
&&B_2=\rho/24,\qquad B_1= 8B_2,\nonumber \\
&&C_2=-\frac{\rho}{24},\qquad C_1=2C_2,
\qquad C_0=-\frac{2\rho}{3},\nonumber \\
&&D_2=\frac{\rho}{16}, \qquad D_1=8D_2\nonumber \\
&&E_{2\alpha\beta}=-\frac{1}{16}(\Pi_{\alpha,\beta} -\frac{\sigma_{\gamma\gamma}}
{3} \delta_{\alpha\beta}),\quad E_{1\alpha\beta}=
8E_{2\alpha\beta}\nonumber \\
&&P_2=\frac{1}{24}, \qquad P_1=8P_2. \label{coeff}
\end{eqnarray}

\noindent{
Clearly the stress tensor $\Pi_{\alpha\beta}$ is a function of
$Q_{\alpha\beta}$ and, in the hybrid approach,  the coefficients
$E_{2\alpha\beta}$ and  $E_{1\alpha\beta}$ of the $f_i^{eq}$'s are
computed by using the solution (via finite difference methods) of
the coupled Eq. (\ref{Qevolution}). This differs from the fully LB
treatment of liquid crystals \cite{Denniston01,Denniston04}.}

\subsection{Rendering of the optical pattern}

\noindent Due to their anisotropic structure, liquid crystal molecules are optically active and cause polarisation rotations 
and phase-shifts in the electric field components of transmitted polarised light.
A common experimental technique is to observe the transmission pattern under a microscope using a crossed-polariser geometry.
Every molecule along the path of the incident beam acts as a retarder. 
Thus the transmission pattern can be regarded as an overall effect of the constituting molecules. 
It depends on the local orientation of the molecules, i.e. on the director field, the refractive-index anisotropy and on the shape of the domain.

While it is in general not trivial to infer 
the local director field from the transmittion pattern, 
it is rather simple to simulate the latter from a given director field. 
 To this end, we have employed the M\"uller matrix 
technique \cite{Berggren94,Bohren}, which simulates the light transmission signal
observed in a micron-size sample under a pair of crossed
polarisers.
We calculated the polarised optical texture corresponding to the 
instantaneous director field of the growing BP domain and its surrounding
environment.In our approach the director field $\vec{d}(\vec{r})$ is defined as the normalized eigenvector 
related to the largest eigenvalue of the order parameter tensor {\bf Q}.
According to Stokes the polarisation state of the light can be conveniently described by a set of four parameters, 
combined in the 4-component Stokes vector $\vec{S}=(S_0,S_1,S_2,S_3)$. 
Its first component $S_0$ is proportional to the intensity of the light.
 
In order to simulate the retardation of a liquid crystal droplet one assumes the droplet of size $L$ 
to consist of $N$ equally thick liquid crystal layers with thickness $h=L/N$.
The effect of each layer situated at the site $\vec{r}$ onto the Stokes-vector $\vec{S}$ is then described by the M\"uller matrix
\begin{equation}
M(\vec{r})=\left(
\begin{array}{cccc}
1 & 0 & 0 & 0 \\
0 & c_b^2+s_b^2 c_d  & c_b s_b(1-c_d) & -s_b s_d\\
0 &  c_b s_b (1-c_d) & s_b^2+c_b^2 c_d & c_b s_d\\
0 & s_b s_d & -c_b s_d & c_d
\end{array} \right),
\end{equation}   
where
\begin{equation}
c_b=\cos{2 \beta(\vec{r})},\, s_b=\sin{2 \beta(\vec{r})},\, c_d=\cos{\delta(\vec{r})},\, s_d=\sin{\delta(\vec{r})}
\end{equation}
and 
\begin{eqnarray}
\alpha(\vec{r})&=&\arccos \left(d_z(\vec{r})\right),\\
\beta(\vec{r})&=&\arctan \left(d_x(\vec{r})/d_y(\vec{r})\right),\\
\delta(\vec{r})&=&\frac{2 \pi q(\vec{r}) h}{\lambda} \left(\frac{n_0 n_e}{\sqrt{n_0^2 \sin^2{\alpha(\vec{r})}+n_e^2\cos^2{\alpha(\vec{r})}}}-n_0\right).
\end{eqnarray}
The quantity $q$ is the local scalar order parameter defined as the largest eigenvalue of the order parameter tensor {\bf Q}, whereas $\lambda$ gives the wavelength of the incident light. 
We set the ordinary and extraordinary refractive indices to $n_0=1.5$ and $n_e=2.0$ respectively.

The above definitions of the angles $\alpha$ and $\beta$ are correct for a light beam propagating along the z-direction. 
We assumed this throughout our analysis, but the adaption to other situations is straightforward.
Generally, $\alpha$ is the angle between the direction of the light beam and the local director field, 
while $\beta$ measures the angle between the projection of the local director field onto the coordinate plane 
perpendicular to the beam direction and a coordinate axis in that plane, in our case the x-axis.
The crossed-polariser geometry is realized by two different M\"uller matrices of the type
\begin{equation}
P=\frac{1}{2} \left(
\begin{array}{cccc}
1 & \cos{\phi} & \sin{\phi} & 0 \\
\cos{\phi} & \cos^2{\phi} & \cos{\phi}\sin{\phi} & 0\\
\sin{\phi} & \sin{\phi}\cos{\phi} & \sin^2{\phi} & 0\\
0 & 0 & 0 & 0
\end{array} \right).
\end{equation}
The parameter $\phi$ defines the angle between the polariser or analyzer and a coordinate axis perpendicular to the beam direction. 
For a right angled crossed-polariser setup one assumes for instance $\phi_{in}=0$ for the polariser and $\phi_{out}=\pi/2$ for the analyser.

The total effect of the liquid crystal droplet on a Stokes vector $\vec{S}_{in}$ can be formally expressed by a matrix product of
consecutive M\"uller matrices, following the path of the light beam:
\begin{equation}
\vec{S}_{out}(x,y)=P_{out}\;\Pi_{k=0}^N\, M(i=x/h,j=y/h,k)\; P_{in} \vec{S}_{in}(x,y).
\end{equation} 
The matrices $M(i,j,k)$ shall be understood as the discretised version of $M(\vec{r})$. 
As we are interested in the general appearance of the sample under polychromatic polarised light, we performed 
this operation for slightly different wavelengths in the magnitude of the 
unit cell size of the BPs, which was $16$ in lattice units, and weighted the results accordingly \cite{Ondris-Crawford90}. We simulated three different components with wavelengths $\lambda=16, 18$ and $20$ in lattice units and assigned them the weights $w=0.2, 0.6$ and $0.2$ respectively. This procedure can only be regarded as a simple approach to model white light. Nevertheless it proves to be sufficient as the results for different components with similar wavelengths differ only slightly in their brightness and only insignificantly in their general transmission patterns. The results for the first component $S_0$ of the Stokes vector $\vec{S}$ are displayed in grayscale in Fig. \ref{fig4}.

\section{Results}

\noindent We now discuss the results obtained by applying our hybrid lattice Boltzmann
algorithm to study the growth of blue phase domains inside cholesteric
liquid crystals. To better analyse the growth dynamics, we restricted
ourselves to the case of a thin slab of liquid crystal, and assumed 
periodic boundary conditions along the small direction, which we took
along the $z$ axis. This corresponds to assuming that the domain wall
is locally straight and should approximate well the experimental
situation in which the growing blue phase droplets are very large with
respect to the unit cell size, so that locally they can be viewed as 
basically planar. In this geometry, which
only requires few (16-32) lattice points along the $z$ direction, we 
can study the time evolution of the order parameter and disclination
network for timescale of up to seconds. In physical units,
the size of one unit cell is about 0.5 $\mu$m.

\begin{figure}[h]
\centerline{\includegraphics[width=7.5in]
{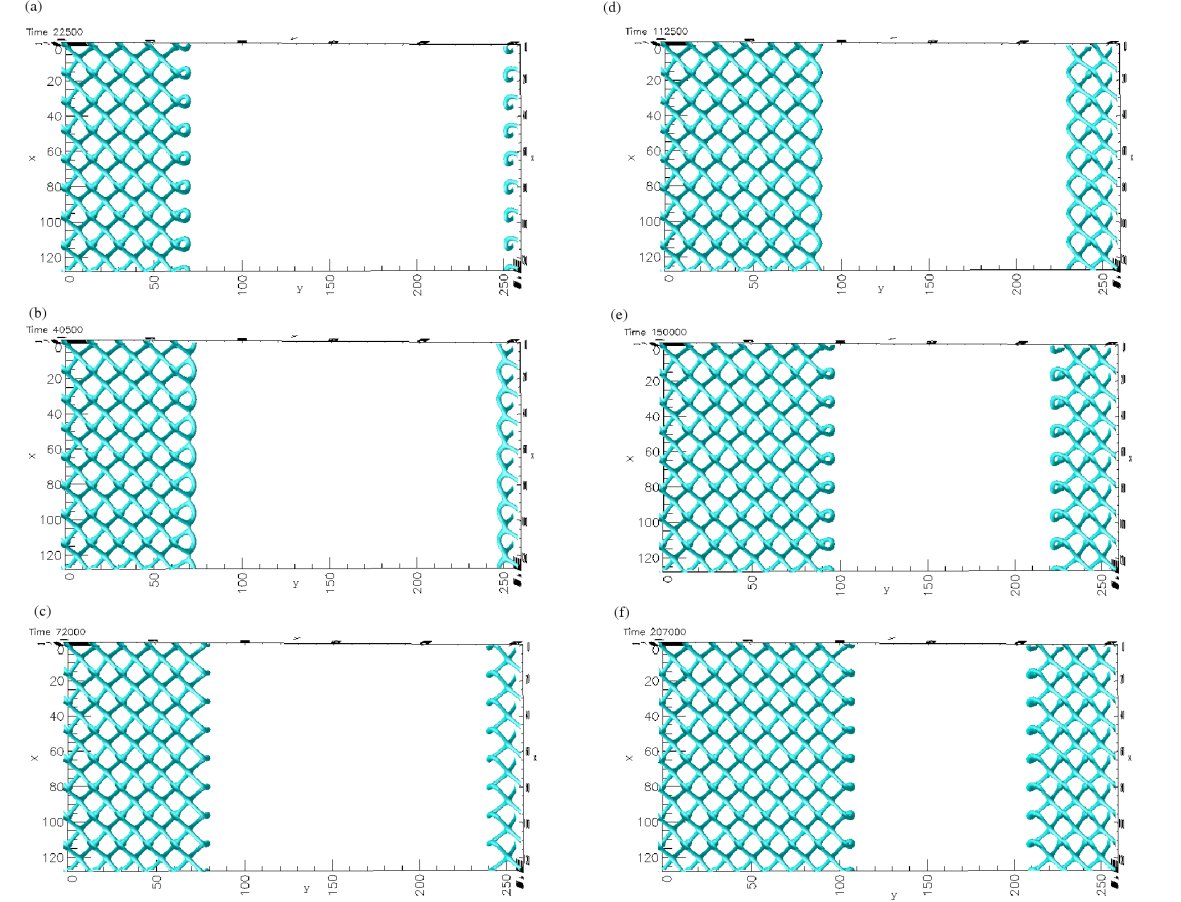}}
\caption{(a)-(f) Snapshots of the 2D projection of the disclination network
of a growing BPII domain inside a cholesteric phase.
Times (in simulation units) are shown in each of the panels. 
The chirality and reduced temperature were $\kappa=1$ and $\tau=0$
respectively.}
\label{fig1}
\end{figure}

In our simulations, a fraction (typically) of the simulated lattice
was initialised in a BPII structure, previously equilibrated at the selected 
values of $\kappa$ and $\tau$, whereas the rest of the lattice was initialised 
in the cholesteric phase. We first consider the case of low chirality
and of a BPII domain growing inside a cholesteric phase 
($\kappa=1$, Fig. \ref{fig1}). 
It can be seen that the growth proceeds
in a regular way: the BPII structure grows, as one would expect as its
free energy is lower than that of the cholesteric (we are in a region in
parameter space in which the cubic BPs are more advantageous structure than
the cholesteric phase). The disclinations twist at the domain boundary to
then merge into regular arrays of BPII unit cells. Their sizes are initially
slightly larger than that of the initial template, but they rapidly 
equilibrate during growth to yield a virtually defect-free BPII lattice 
at the end of the simulations. 

\begin{figure}[h]
\centerline{
\includegraphics[width=7.5in]
{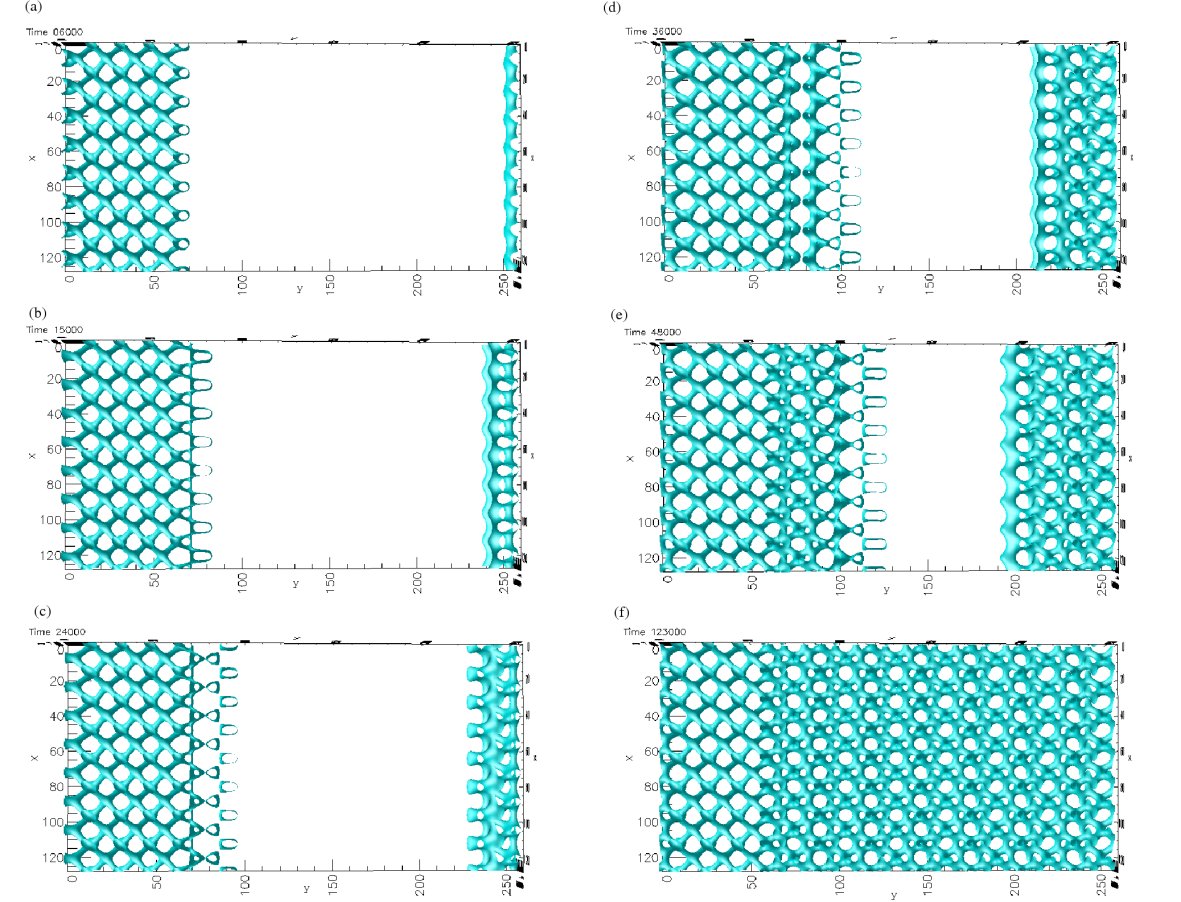}}
\caption{(a)-(f) Snapshots of the 2D projection of the disclination network
of a growing BPII domain inside a cholesteric phase.
Times (in simulation units) are shown in each of the panels. 
The chirality and reduced temperature were $\kappa=2$ and $\tau=0$
respectively.}
\label{fig2}
\end{figure}

Increasing the chirality (Fig. \ref{fig2}), we observe 
an interesting phenomenon. The growth is much faster, and is
more irregular at the surface. The 
twist at the advancing boundary can no longer be
accomodated within the BP II lattice. Therefore, for large
enough values of the chirality, it becomes more advantageous
to switch to a completely different blue phase lattice, with hexagonal 
structure. This hexagonally ordered structure is different from
the previously reported hexagonal phases, which were theoretically predicted 
to be stable only in
weak electric fields \cite{Hornreich85}. The BP we report here
is not, so far as we know, observed experimentally,
but it would be interesting to understand whether it could be related
to local ordering in blue phase III, which appear at a larger chirality
than either BPI or BPII. It is quite possible anyway for such a phase
to arise as a metastable intermediate.

The domain wall growth is very anisotropic. The results we have 
presented are with the cholesteric helix pointing along the $y$ axis,
perpendicular to the domain wall. We have performed additional 
simulations with different relative orientation between helix and
domain wall. If the helix lies on the plane of the domain wall,
we do not observe the transition to the hexagonally ordered structure.
The anisotropy is also clear in the orientation we have chosen
(Figs. \ref{fig1} and \ref{fig2}),
as the speed of advance of the domain boundary is not equal at both ends.

\begin{figure}[h]
\centerline{
\includegraphics[width=7.5in]
{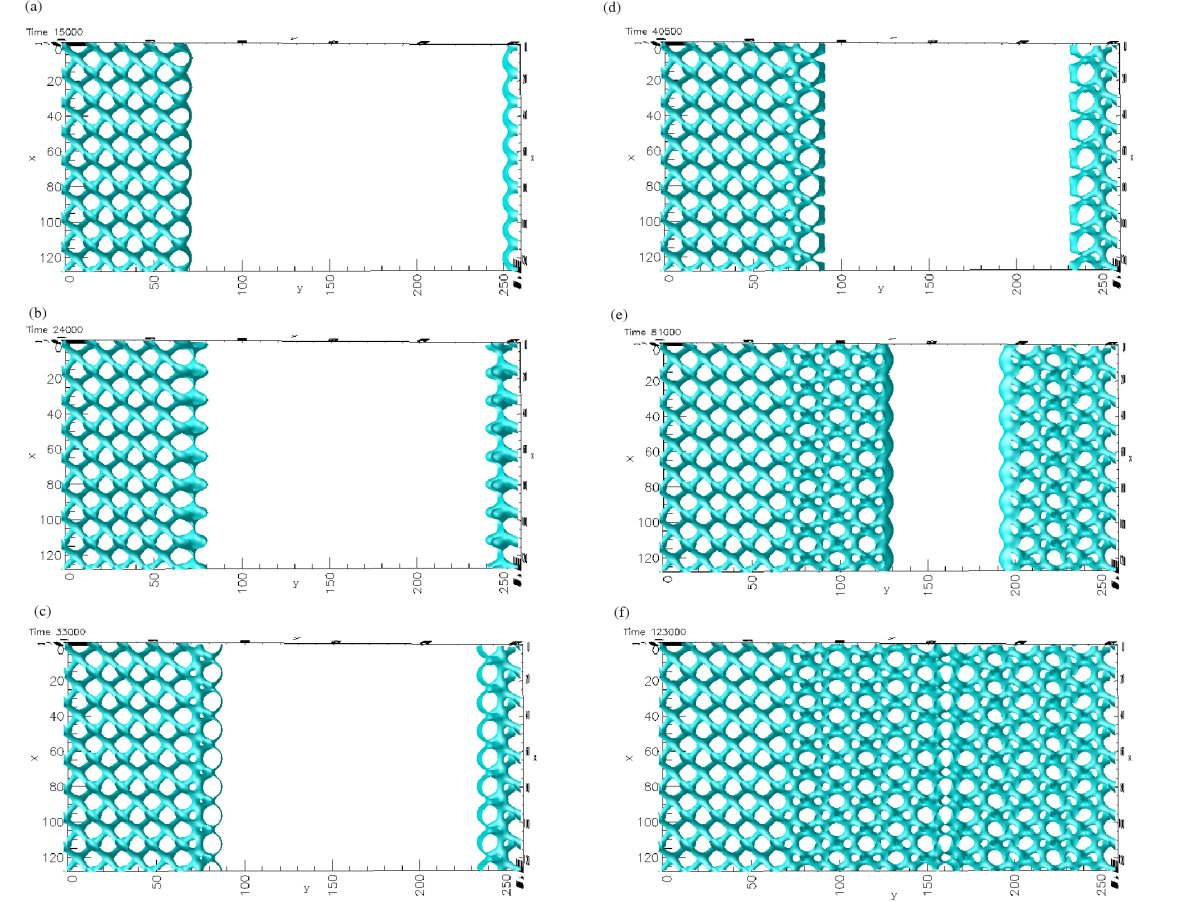}}
\caption{(a)-(f) Snapshots of the 2D projection of the disclination network
of a growing BPII domain inside an isotropic phase.
Times (in simulation units) are shown in each of the panels. 
The chirality and reduced temperature were $\kappa=2$ and $\tau=0$
respectively.}
\label{fig3}
\end{figure}

We have also performed simulations in which BPII was growing inside
an isotropic phase at the same values of the reduced temperature.
Again we find evidence of a transition between different BP lattices.
At high values of the chiralities the
``hexagonal'' BP again forms, albeit through a different kinetic mechanism:
instead of forming twisted disclination loops which are joined later
on as within the cholesteric phase, the structure extends more
regularly (Fig. \ref{fig3}). This reinforces the 
suggestion that at high chirality the hexagonal ordering
of the BP lattice should be thermodynamically advantageous.

\begin{figure}[h]
\centerline{
\includegraphics[width=7.5in]
{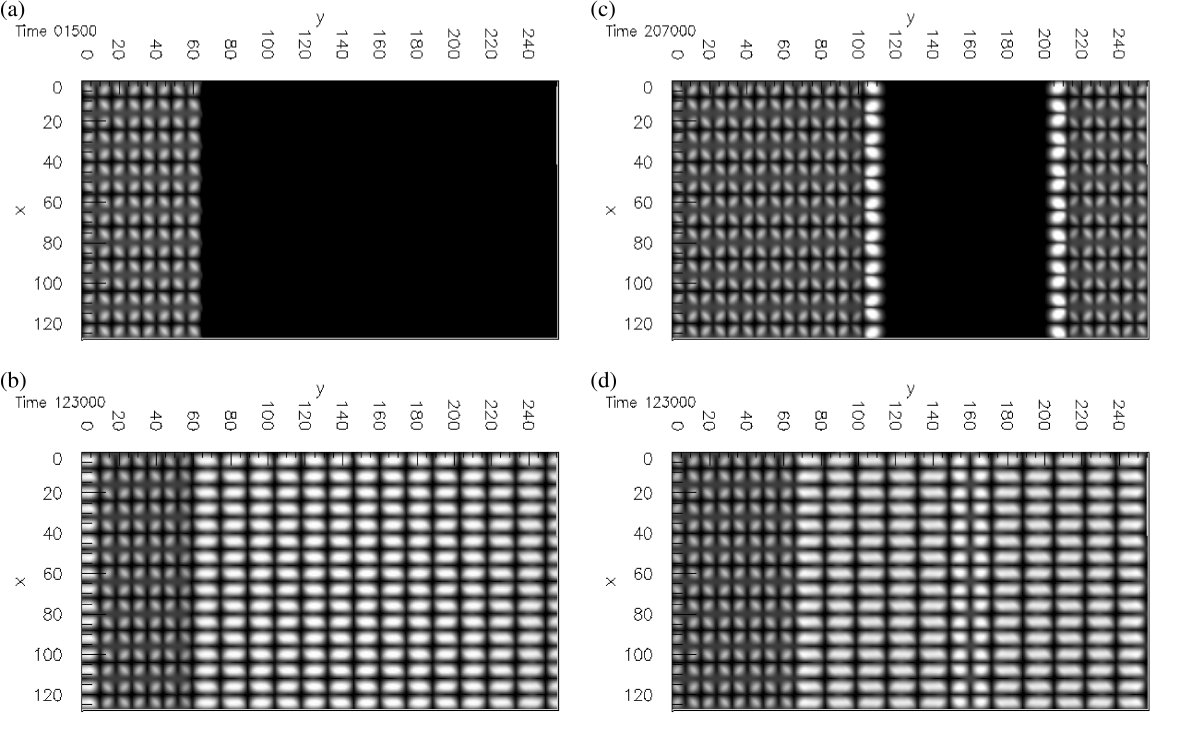}}
\caption{Predicted optical patters for the (a) initial configuration,
and final simulated configurations for (b) $\kappa=1$ and a BPII in
an initially cholesteric phase; (c) $\kappa=2$ and a BPII in 
an initially 
cholesteric phase; (d) $\kappa=2$ and a BPII in an initially isotropic}
\label{fig4}
\end{figure}

Finally, while Figs. \ref{fig1}, \ref{fig2}, \ref{fig3} show the time 
evolution of the disclination line network, Fig. 
\ref{fig4} gives the corresponding predicted optical
patterns (one should keep in mind that the achievable resolution in 
an experiment would be smaller). It can be seen that the hexagonally
ordered phase leaves a distinct signature from the BPII structure.
Note that the hexagonal ordering is particularly evident in the 
larger corner spots in the optical pattern as well as generally 
in the $\sqrt{3}:1$ aspect ratio of the rectangular pattern.

\section{Conclusions}

\noindent In conclusion, we have presented a hybrid lattice Boltzmann algorithm to solve 
the Beris-Edwards hydrodynamic equations of motion of cholesteric blue phases.
The hybrid methodology mainly allows us to decrease the memory requirements, 
which in previous full LB approaches eventually limited the lattice sizes
which could be reached.

We have shown that it is possible with this algorithm to perform supradomain
simulations of blue phases, in which a large number of unit cells is
simultaneously considered, as opposed to the previously reported single
unit cell simulations, which assumed a regular lattice of disclination
lines. In particular we have reported results on the growth of blue phase II
domains in cholesterics, and shown that the growth dynamics is highly
non-trivial. For low values of the chirality, the growth is regular and
the blue phase forms without defects, whereas for large values of the 
chirality the growing region no longer has the structure of blue phase II, 
but attains a hexagonal ordering, and several defects appear at late times
when the blue phase structure has invaded the whole computational domain.
Finally, for very large values of the chirality, we observe a more regular
hexagonal blue phases at late times. The switch between blue phase II and
this hexagonal structure is due, in our interpretation, to the increasing
twist which is developed at the domain wall, and which can no longer relax
into a regular blue phase II unit cell. Our predicted transition
from blue phase II to hexagonal structure appears for rather large 
values of the chirality, and it would be interesting to see whether these
could be reached by the new temperature-stabilised blue phases.
Also, it would be of interest to see whether this transition can
be explained with extensions of the existing semi-analytical theories
of blue phases, which are based on truncated expansions of the order parameter
in Fourier components.

We acknowledge EPSRC grant EP/E045316/1 for funding, and. G. P. Alexander
and E. Orlandini for useful discussions and cpu time on
Hector funded by EP/F054750/1.


\begin{thebibliography}{99}
\bibitem{Succi} S. Succi, {\it The Lattice Boltzmann Equation for Fluid 
Dynamics and Beyond}, Oxford University Press (2001).
\bibitem{Boghosian} B. M. Boghosian, {\it Computing in Science and 
Engineering}, {\bf 5},  86 (2003).
\bibitem{bijel} K. Stratford, R. Adhikari, I. Pagonabarraga, J.-C. Desplat 
and M. E. Cates, {\it Science} {\bf 309}, 2198 (2005).
\bibitem{bijel_review} P. S. Clegg, {\it J. Phys.: Condens. Matt.}
{\bf 20}, 113101 (2008).
\bibitem{Swift96} M. R. Swift, E. Orlandini, W. R. Osborn and J. M. Yeomans,
{\it Phys. Rev. E} {\bf 54}, 5041 (1996).
\bibitem{Wagner98} A. J. Wagner and J. M. Yeomans, {\it Phys. Rev. Lett.}
{\bf 80}, 1429 (1998).
\bibitem{Gonnella97} G. Gonnella, E. Orlandini and J. M. Yeomans, 
{\it Phys. Rev. Lett.} {\bf 78}, 1695 (1997).
\bibitem{Xu06} A. G. Xu, G. Gonnella and A. Lamura,
{\it Phys. Rev. E} {\bf 74}, 011505 (2006).
\bibitem{Xu06HLB} A. G. Xu, G. Gonnella and A. Lamura, 
{\it Physica A} {\bf 362}, 42 (2006). 
\bibitem{Denniston04} C. Denniston,
D. Marenduzzo, E. Orlandini and J. M. Yeomans, 
{\it Phil. Trans. R. Soc. Lond. A} {\bf 362}, 1745 (2004).
\bibitem{Marenduzzo07} D. Marenduzzo, 
E. Orlandini, M. E. Cates and J. M. Yeomans, 
{\it Phys. Rev. E} {\bf 76}, 031921 (2007).
\bibitem{Cates08} M. E. Cates, S. M. Fielding, D. Marenduzzo, E. Orlandini
and J. M. Yeomans, {\it Phys. Rev. Lett.} {\bf 101}, 068102 (2008).
\bibitem{Kevin1} K. Stratford, J.-C. Desplat, P. Stansell and M. E. Cates,
{\it Phys. Rev. E} {\bf 76}, 030501 (2007).
\bibitem{Kevin2} P. Stansell, K. Stratford, J.-C. Desplat, R. Adhikari and
M. E. Cates, {\it Phys. Rev. Lett.} {\bf 96}, 085701 (2006).
\bibitem{Beris} A. N. Beris and B. J. Edwards, {\it Thermodynamics of Flowing
Systems}, Oxford Science Publications (1994).
\bibitem{Ronojoy} R. Adhikari and I. Pagonabarraga, paper in preparation. 
\bibitem{Wright89} D. C. Wright and N. D. Mermin, 
{\it Rev. Mod. Phys.} {\bf 61}, 385 (1989).
\bibitem{Coles05} H. J. Coles and M. N. Pivnenko, {\it Nature} {\bf 436}, 
977 (2005).
\bibitem{Pieranski85} P. Pieranski, R. Barbet-Massin and P. E. Cladis,
{\it Phys. Rev. A} {\bf 31}, 3912 (1985).
\bibitem{Pieranski00} P. Pieranski, {\it Phys. Rev. Lett.} {\bf 84}, 
2409 (2000).
\bibitem{Pieranski08} P. Pieranski, {\it Europhys. Lett.} {\bf 81}, 66001 
(2008).
\bibitem{Dupuis05a} A. Dupuis, D. Marenduzzo, and J. M. Yeomans, 
{\it Phys. Rev. E} {\bf 71}, 011703 (2005).
\bibitem{Alexander06} G. P. Alexander and J. M. Yeomans, {\it Phys. Rev. E} 
{\bf 74}, 061706 (2006). 
\bibitem{Dupuis05b} A. Dupuis, D. Marenduzzo, E. Orlandini, and J. M. Yeomans, 
{\it Phys. Rev. Lett.} {\bf 95}, 097801 (2005).
\bibitem{Alexander08} G. P. Alexander and D. Marenduzzo, {\it Europhys. Lett.}
{\bf 81}, 66004 (2008).
\bibitem{Grebel} H. Grebel, R. M. Hornreich and S. Shtrickman, 
{\it Phys. Rev. A} {\bf 30}, 3264 (1984).
\bibitem{Larson} R. G. Larson, {\it The Structure and Rheology of
Complex Fluids}, Oxford University Press, USA (1999).
\bibitem{Denniston01} C. Denniston, E. Orlandini, J. M. Yeomans, 
{\it Phys. Rev. E}  {\bf 63}, 056702 (2001).
\bibitem{LBnoise} R. Adhikari, K. Stratford, M. E. Cates and A. J. Wagner,
{\it Europhys. Lett.} {\bf 71}, 473 (2005).
\bibitem{note_bodyforce} 
We note that in principle when entering a 
body force, we should add a correction term in the second moment of
the $p_i$'s, see for instance A. J. Wagner, {\it Phys. Rev. E} 
{\bf 74}, 056703 (2006). This may have a quantitative effect, 
although theoretically in our case the correction
terms in the Chapman-Enskog expansion would be third order in gradients, 
and these are commonly neglected~\cite{Denniston01}.
\bibitem{Wagner03} A. J. Wagner,
{\it Int. J. Mod. Phys. B} {\bf 17}, 193 (2003).
\bibitem{Berggren94} E. Berggren, 
C. Zannoni, C. Chicolli, P. Pasini and F. Semeria, 
{\it Phys. Rev. E} {\bf 50}, 2929 (1994).
\bibitem{Bohren} C.F. Bohren and D.R. Huffman, 
{\it Absorption and Scattering of Light by Small Particles}, 
Wiley, New York (1998).
\bibitem{Ondris-Crawford90} R. Ondris-Crawford, E.P. Boyko, B.G. Wagner, J.H. Erdmann, {\it J. Appl. Phys.} {\bf 69}, 6380 (1991).
\bibitem{Hornreich85} R.M. Hornreich, M. Kugler, S. Shtrikman, {\it Phys. Rev. Lett.} {\bf 54}, 2009 (1985).

\end{thebibliography}
\end{document}